\documentclass[onecolumn]{emulateapj}
\usepackage{amsmath}

\lefthead{Wada, Norman}
\righthead{log-normal pdf}

\newcommand{\bea}{\begin{eqnarray} }
\newcommand{\eea}{\end{eqnarray}}
\def\mbf#1{\mbox{\boldmath ${#1}$}}

\newcommand{\arho}{\rho_0}

\begin{document}	

\title{Density structure of the interstellar medium and the star formation rate in galactic disks}

\author{Keiichi Wada$^1$}
\affil{National Astronomical Observatory of Japan, Mitaka, Tokyo
181-8588, Japan\\
E-mail: wada.keiichi@nao.ac.jp}
\altaffiltext{1}{Department of Astronomical Science, The Graduate University for Advanced Studies, Osawa 2-21-1, Mitaka, Tokyo 181-8588, Japan.}

\author{Colin A. Norman$^2$}
\affil{Johns Hopkins University, Baltimore, MD 21218\\
E-mail: norman@pha.jhu.edu}
\altaffiltext{2}{Space Telescope Science Institute, Baltimore, MD 21218}



\begin{abstract}
The probability distribution functions (PDF) of density of the
interstellar medium (ISM) in galactic disks and global star formation
rate are discussed.  Three-dimensional hydrodynamic simulations show
that the PDFs in globally stable, inhomogeneous ISM in galactic disks
are well fitted by a single log-normal function over a wide density
range. The dispersion of the log-normal PDF (LN-PDF) is larger for
more gas-rich systems, and whereas the characteristic density of
LN-PDF, for which the volume fraction becomes the maximum, does not
significantly depend on the initial conditions.
Supposing the galactic ISM is characterized by the LN-PDF, we give a
global star formation rate (SFR) as a function of average gas density,
a critical local density for star formation, and star formation
efficiency. Although the present model is more
appropriate for massive and geometrically thin disks ($\sim 10$
pc) in inner galactic regions ($ <$ a few kpc), we can make a
comparison between our model and observations in terms of SFR,
provided that the log-normal nature of the density field is also the
case in the real galactic disk with a large scale height ($\sim 100$
pc).  We find that the observed SFR is well-fitted by the theoretical
SFR in a wide range of the global gas density ($10 - 10^4 M_\odot$
pc$^{-2}$).  Star formation efficiency (SFE) for high density gas ($n
> 10^3$ cm$^{-3}$) is SFE $= 0.001 - 0.01$ for normal spiral galaxies,
and SFE $= 0.01 - 0.1$ for starburst galaxies.
The LN-PDF and SFR proposed here could
be applicable for modeling star formation on a kpc-scale in galaxies
or numerical simulations of galaxy formation, in which the numerical
resolution is not fine enough to describe the local star formation.

\end{abstract}




%


%
\section{INTRODUCTION}
%
The interstellar medium (ISM) in galaxies 
is characterized by highly inhomogeneous structure
with a wide variety of physical and chemical states \citep{myer78}. 
Stars are formed in this complexity through gravitational instability 
in molecular cores, 
but the entire multi-phase structures on a global scale are quasi-stable.
It is therefore important to understand theoretically
the structure of the ISM in a wide dynamic range to
model star formation in galaxies. 
Observations suggest that there is a positive correlation between global star formation
rate (SFR) and the average gas density:
$ \dot{\Sigma}_{\star} \propto \Sigma_{gas}^N $ with
$N \sim 1.4$ in nearby galaxies \citep{ken98}\footnote{
Recent observations reported a wide variety on the slope, e.g.
$N\sim 1.1$ or 1.7 depending on the extinction models (Wong \& Blitz 2002).
\citet{kom05} found that $N\sim 1.33$ for the central part of normal galaxies, and
they also suggest that SFR is systematically 
lower than those in starburst galaxies (See also \S 3.3).}.
Since star formation process itself is a local phenomenon on a sub-parsec scale,
the observed correlation between the structure of the ISM on a local scale and the
global quantities, such as the average gas density, 
implies that the ISM on different scales is physically related.

In fact, 
 two- and 
three-dimensional hydrodynamic and magneto-hydrodynamic simulations
 \citep[e.g.][]{BL,VZ00,rosen95,avil00} show that
there is a robust relation between the local and global structures of
the multi-phase ISM, which is described by 
a log-normal density PDF (probability distribution function).
Elmegreen (2002) first noticed that if the density 
PDF is log-normal and star formation occurs in dense gases above a
critical density,
the Schmidt-Kennicutt law is reproduced.
This provides a new insight on the origin of the scaling relation.
More recently \citet{krum05} give a similar model on SFR in molecular clouds.
In these theoretical predictions, dispersion of the LN-PDF, $\sigma$, is 
a key parameter. Elmegreen (2002) used $\sigma = 2.4$, which is taken from
2-D hydrodynamic simulations of the ISM \citep{wad01b}. \citet{krum05}
assumed an empirical relation between $\sigma$ and the rms Mach number,
which is suggested in numerical simulations of isothermal turbulence
(see also \S 4.1).
Therefore, it is essential to know whether the LN-PDF 
in galactic dicks is universal, and
what determines $\sigma$.
However, most previous simulations, in which LN-PDF or power-law PDF are reported,
are `local' simulations:  a patch of the galactic disk is 
simulated with a periodic boundary condition.
Apparently, such local simulations are not suitable to
discuss statistical nature of the ISM in galactic disks. For example,
number of density condensations are not large enough (typically a few)
\citep[see][]{scal98, slyz05}.

On the other hand, global hydrodynamic simulations for two-dimensional galactic disks
or three-dimensional circum-nuclear gas disks suggested that 
the density PDF, especially a high-density part
is well fitted by a single log-normal function over 4-5 decades
(Wada \& Norman 1999, 2001, hereafter WN01 and Wada 2001).
The log-normal PDF is also seen in a high-z galaxy formed by 
a cosmological N-body/AMR(Adaptive Mesh Refinement) simulation \citep{krav03}.
Nevertheless universality of LN-PDF and how it is related to
global quantities are still unclear.


In this paper, we verify the log-normal nature of the ISM in galactic 
disks, using three-dimensional, global hydrodynamic simulations.
 This is an extension of
our previous two-dimensional studies of the ISM in galactic disks  (WN01)
 and three-dimensional model in the galactic central region
 \citep[][]{wad02}.
We confirm that dispersion $\sigma$ of the log-normal function
is related to average gas density of the disk.
We then calculate SFR as a function of critical density of local
star formation and star star formation efficiency.
This is a generalized version of the Schmidt-law \citep{sch59}, and 
it can be applied to various situations. 

An alternative way to study the global SFR in galaxies is 
simulating star formation directly in numerical models \citep[e.g.][]{li05, krav05,task06}.  In this approach, `stars' are formed according to a `star formation recipe',
 and the resultant SFR is compared to observations. 
We do not take this methodology, because numerical modeling star formation 
in simulations 
still requires many free-parameters and assumptions.
Moreover if the numerically 
obtained SFR deviates from the observed scaling relation (this is 
usually the case. See for example Tasker \& Bryan 2006), it is hard to
say what we can learn from the results. This deviation might be due to
wrong implementation of star formation in the numerical code,
or estimate of SFR in observations might be wrong, since SFR is not
directly observable.  
 One should also realize that 
comparison with observations of local galaxies is not necessarily 
useful when we discuss SFR in different situations, such as galaxy formation.
In this paper, we avoid
the ambiguity in terms of star formation in simulations,
and alternatively discuss SFR based on an intrinsic statistical feature of the ISM.
Effect of energy feedback from supernovae is complementarily discussed.

This paper is organized as follows. 
In \S 2, we describe results of numerical simulations 
of the ISM in a galactic disk, 
focusing on the PDF.
In \S 3, we start from
a `working-hypothesis', that is 
the inhomogeneous ISM, which is formed through non-linear 
development of density fluctuation, is characterized by 
a log-normal density PDF. 
After summarizing basic properties of the LN-PDF,
we use the LN-PDF to estimate a global star formation rate,
and it is compared with observations.
In \S 4, we discuss implications of 
the results in \S 2 and \S 3.
In order to distinguish local (i.e. sub-pc scale) and global (i.e.
galactic scale) phenomena, we basically use
cm$^{-3}$ for local number density and $M_\odot$ pc$^{-3}$ (or $M_\odot$ pc$
^{-2}$)
for global density\footnote{1$M_\odot$ pc$^{-3} \simeq 66.7$ cm$^{-3}$, if
the mean weight of a particle is 0.61$m_{\rm H}$.}.

%
\section{GLOBAL SIMULATIONS OF THE ISM IN GALACTIC DISKS AND THE PDF}
%
\subsection{Numerical Methods}
Evolution of rotationally supported gas disks in a
fixed (i.e. time-independent), spherical galactic potential
is investigated using three-dimensional hydrodynamic simulations.
We take into account self-gravity of the gas, radiative cooling and heating
processes. The numerical scheme is an Euler method with a uniform Cartesian
grid, which
is based-on the code described in WN01 and \citet{wad01}.
Here we briefly summarize them.
\setcounter{footnote}{0} We solve the following 
conservation equations and Poisson equation
in three dimensions.
  \begin{eqnarray}
\frac{\partial \rho}{\partial t} + \nabla \cdot (\rho \mbf{v}) &=& 0,
\label{eqn: rho} \\ \frac{\partial \mbf{v}}{\partial t} + (\mbf{v}
\cdot \nabla)\mbf{v} +\frac{\nabla p}{\rho} + \nabla \Phi_{\rm ext} +
\nabla \Phi_{\rm sg} &=& 0, \label{eqn: rhov}\\
 \frac{\partial (\rho E)}{\partial t} + \nabla \cdot 
[(\rho E+p)\mbf{v}] &=& 
\rho \Gamma_{\rm UV} + \Gamma_{\rm SN} - \rho^2 \Lambda(T_g), \label{eqn: en}\\ \nabla^2
\Phi_{\rm sg} &=& 4 \pi G \rho, \label{eqn: poi} 
\end{eqnarray}

 where, $\rho,p,\mbf{v}$ are the density, pressure, velocity of
the gas. The specific total energy $E \equiv |\mbf{v}|^2/2+
p/(\gamma -1)\rho$, with $\gamma= 5/3$.  
The spherical potential is $\Phi_{\rm ext} \equiv  -(27/4)^{1/2} [v_1^2/(r^2+
a_1^2)^{1/2} + v_2^2/(r^2+a_2^2)^{1/2}]$, where $a_1 = 0.3$ kpc, 
$a_2 = 5$ kpc, and $v_1 = v_2 = 200$ km s$^{-1}$.
 We also assume a cooling function
$\Lambda(T_g) $ $(10 < T_g < 10^8 {\rm K})$ \citep{spn97} with solar metallicity.
We assume photo-electric heating by dust and a uniform UV radiation field,
$ \Gamma_{\rm UV} = 1.0 \times 10^{-23} \varepsilon   
G_0 \, {\rm ergs \: s}^{-1}, $
where the heating efficiency $\varepsilon$ is assumed to be 0.05 and
$G_0$ is the incident FUV field normalized to the local interstellar
value \citep{ger97}.
In order to focus on a intrinsic inhomogeneity in the ISM due to 
gravitational and thermal instability, 
first we do not include energy feedback from supernovae.
In \S 4.5 we show a model with energy feedback from supernovae (i.e. $\Gamma_{\rm SN} \ne 0$).
Note that even if there is no random energy input 
from supernovae, turbulent motion can be
maintained in the multi-phase, inhomogeneous gas disk \citep{wad02b}.

The hydrodynamic part of the basic equations is solved by AUSM
 (Advection Upstream Splitting Method) \citep{LS} with a
uniform Cartesian grid. 
We achieve third-order spatial accuracy with MUSCL (Monotone  Upstream-centered Schemes for Conservation Laws) \citep{VL}. 
We use $512\times 512 \times 64$ grid points covering a $2.56\times 2.56\times 0.32$ kpc$^3$ region (i.e. the spatial resolution is 5 pc).
For comparison, we also run models with a 10 pc resolution.
The Poisson equation is solved to calculate self-gravity of the gas
using the Fast Fourier Transform (FFT)
 and the convolution method \citep{hock81}. 
In order to calculate the isolated gravitational potential of the gas, 
FFT is performed for a working region of $1024\times 1024 \times 128$ 
grid points (see details in WN01).
We adopt implicit time integration for the cooling term.
We set the minimum temperature, for which the Jeans instability 
can be resolved with a grid size $\Delta$, namely
$T_{\rm min} = 35$ K ($\rho/10 M_\odot$ pc$^{-3})(\Delta/5 \; {\rm pc})^2$.

The initial condition is an axisymmetric and rotationally supported
thin disk (scale height is 10 pc) with a uniform density $\rho_i$.
We run four models with different $\rho_i$ (see Table 1).
Random density and temperature fluctuations are added to the
initial disk. These fluctuations are less than 1 \% of the unperturbed
values and have an approximately white noise distribution. The initial
temperature is set to $10^4$ K over the whole region. 
At the boundaries, all physical quantities remain at their initial
values during the calculations. 


\begin{table}
\begin{center}
\caption{Initial density and fitting parameters of the density PDF\label{tbl-2}}
\begin{tabular}{lcccccccc}
\tableline\tableline
Model &
\multicolumn{1}{c}{$\rho_i$\tablenotemark{a}}($M_\odot$ pc$^{-3}$) &
\multicolumn{1}{c}{$\sigma_{10}$\tablenotemark{b}} &
\multicolumn{1}{c}{$\sigma$\tablenotemark{c}} &
\multicolumn{1}{c}{$\log \rho_0$\tablenotemark{d}} &
\multicolumn{1}{c}{$\alpha$\tablenotemark{e}} &
\multicolumn{1}{c}{$\sigma_{10,M}$\tablenotemark{f}} &
\multicolumn{1}{c}{$\sigma_M$\tablenotemark{g}} &
\multicolumn{1}{c}{$\sigma_p$\tablenotemark{h}} \\
\tableline 
A  & 5  & 1.025 &2.360 &-1.40 &0.09& 1.025 & 2.360 & 2.483\\
B  & 10 & 1.188 &2.735 &-1.55&0.12 & 1.188 & 2.735 & 2.769\\
C  & 15 & 1.223 &2.816 &-1.60 &0.18& 1.273 & 2.849 & 2.810\\
D  & 50 & 1.308 &3.012 &-1.50 &0.20& 1.308 & 3.012 & 3.104\\
\tableline
\end{tabular}
\tablenotetext{a}{initial density}
\tablenotetext{b}{Dispersion of volume-weighted LN-PDF (eq. (\ref{eqn: fln}))}
\tablenotetext{c}{$\sigma \equiv \sigma_{10} \ln 10$}
\tablenotetext{d}{The reference density for the volume-weighted PDF (eq. (\ref{eqn: fln})).
Unit of density is $M_\odot$ pc$^{-3}$.}
\tablenotetext{e}{Volume fraction of the log-normal part (eq. (\ref{eqn: fln}))}
\tablenotetext{f}{Dispersion of mass weighted LN-PDF (Fig. \ref{fig: fmass}). }
\tablenotetext{g}{$\sigma_M \equiv \sigma_{10,M} \ln 10$. If the PDF is
a perfect log-normal, $\sigma = \sigma_M$}
\tablenotetext{h}{Dispersion of LN-PDF predicted from eq. (\ref{eqn: eq17c}).}
\end{center}
\end{table}
%
\subsection{Numerical Results}
\begin{figure}[t]
\centering
\includegraphics[width = 8cm]{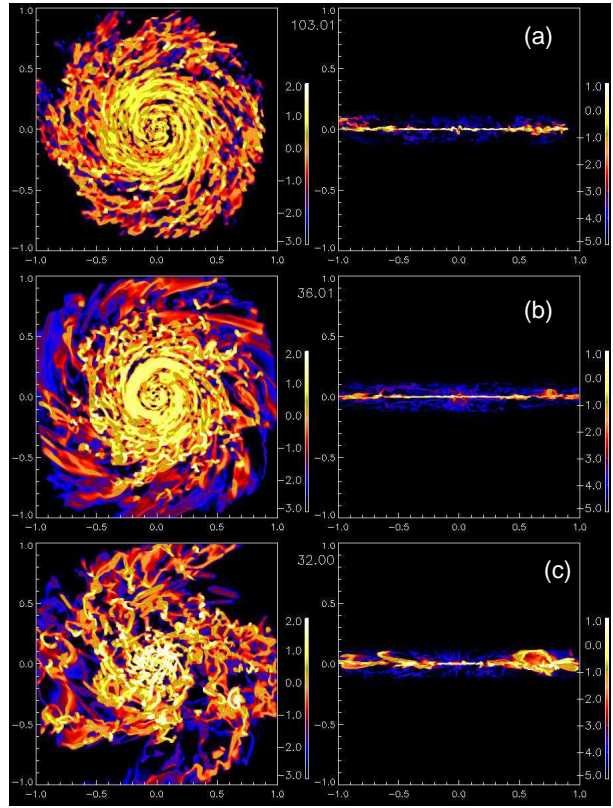}
\caption{Surface sections of the density distribution of the gas in 
models with different initial density (model A, B, and D).  x-y and x-z planes are shown. Unit of length is kpc. (a) model A ($\rho_i = 5 M_\odot$ pc$^{-3}$, $t=103$ Myr), (b) model B ($\rho_i = 10 M_\odot$ pc$^{-3}$, $t=36$ Myr), and (c) model D ($\rho_i = 50 M_\odot$ pc$^{-3}$, $t=32$ Myr).}
\label{fig: f1}
\end{figure}

%
Figure \ref{fig: f1} is
snapshots of density distribution at a quasi-stable state for
models A, B and D on the x-y plane and x-z plane.
Depending on the initial density 
($\rho_i = 5, 10$, and 50 $M_\odot$ pc$^{-3}$), 
distribution of the gas in a quasi-equilibrium is different.
In the most massive disk (Fig. 1c, model D), the disk is fragmented into 
clumps and filaments, on the other hand,
 the less massive disk (Fig. 1a, model A)
shows more axi-symmetric distribution with tightly winding spirals and
filaments.
In Fig. \ref{fig: fig_vertical}, we show vertical structures 
of density of model D ($t=42$ Myr). The high density disk ($\rho > 0.1 M_\odot$ pc$^{-3}$) is resolved by about 10 grid points out of 64 total grid points for
$z$-direction. Density change in about 5 orders of magnitude from the disk
to halo gas is resolved.

Figure \ref{fig: f2} is time evolution of a density PDF in model D.
The initial uniform density distribution, which forms the peak around
$\rho \simeq 50 M_\odot$ pc$^{-3}$, is smoothed out in $\sim 10$ Myr,
 and it turns out to a smooth distribution in $\sim 20$ Myr. 
Figure \ref{fig: f3a} is PDFs spatially de-convolved to
four components (thin disk, 
thick disk in the inner disk, halo, and the whole computational box. See 
figure caption for definitions) 
of model D in a quasi-steady state ($t=32$ Myr).
It is clear that dense regions ($\rho \gtrsim 0.01 M_\odot$
pc$^{-3}$) can be fitted by a single log-normal function, 
$f_{\rm LN}$, over nearly 6 decades:
\bea
f_{\rm LN}(\rho; \arho, \sigma_{10}) = \frac{\alpha}{\sqrt{2\pi}\sigma_{10}\ln 10} 
\exp{\left[
-\frac{
\log (\rho/\arho)^2
}{
2\sigma_{10}^2
}
\right] },
\label{eqn: fln}
\eea
where $\alpha$ is volume fraction of
the high-density part which is fitted by the LN-PDF. 
On a galactic scale, the ISM is a multi-phase, and dense, cold gases 
occupy smaller volumes than diffuse, hot gases.
Therefore, it is reasonable that the density PDF shows a negative 
slope against density as shown in the previous numerical simulations.
However, our results suggest that
structure of the ISM is not scale-free.
The LN-PDF implies that formation process for the 
high density part is highly non-linear \citep{vazq94}.
High density regions can be formed by convergent processes, such as
mergers or collisions between clumps/filaments, compressions by
sound waves or shock waves.
Tidal interaction between clumps and galactic shear or 
local turbulent motion can also change their density structure.
For a large enough volume, and  for a long enough period, 
these processes can be regarded as many random, 
independent processes in a galactic disk. 
In this situation, the density in a small volume is determined by
a large number of independent random events, which can be
expressed by $\rho = \Pi_{i=1}^{i=N} \delta_i \rho_s$ with
independent random factors $\delta_i$ and initial density $\rho_s$.
Therefore the distribution function of $\log(\rho)$ should be
Gaussian according to the central limit theorem, if $N \rightarrow \infty$.

In Table 1, we summarize fitting results for the PDF 
in the four models in terms of the whole volume.
One of important results is that there is a clear trend 
that the dispersion is larger for more massive
systems.
 The less massive model (model B) also 
shows a LN-PDF (Fig. \ref{fig: f3b}),
but the dispersion is smaller ($\sigma_{10} = 1.188$) than that in model D.
The PDF for the whole volume is log-normal, but 
there is a peak around $\log \rho = 1.3$.
As shown by the dotted line,
this peak comes from the gas just above the disk plane 
($10 \leq  z \leq 20$ pc) in the inner disk
($r \lesssim 0.38$ kpc), where the gas is dynamically stable, 
and therefore the density is not very 
different from its initial value ($\rho_i = 10$).
This peak is not seen in the dotted line in model D (Fig. \ref{fig: f3a}), 
since the density field in model D is
not uniform even in the inner region due to the high 
initial density.

The excess of the volume at low density ($\lesssim 10^{-3} M_\odot$ pc$^{-3}$,
see Figs. \ref{fig: f3a} and \ref{fig: f3b}) is 
due to a smooth component that is extened vertically outside the thin, dense 
disk.  About 10-20\% of the whole computational box is in the LN regime (i.e. $\alpha= 0.1-0.2$). However, most fraction of mass is in the LN regime
(see Fig. \ref{fig: fmass}).


In order to ensure that gravitational instability in
high density gas is resolved, 
we set the minimum temperature $T_{\rm min}$, depending on
a grid-size (\S 2.1). 
We run model C with two different resolutions, 5 pc and 10 pc.
As seen in Fig. \ref{fig: f4}, although the ``tangled web'' structure 
is qualitatively similar between the two models, 
typical scales of the inhomogeneity, e.g. 
width of the filaments, are different.
The scale height of the disk $H$ in the model
with $\Delta = 5$ pc is about a factor of two smaller than
the model with $\Delta = 10$ pc. 
This is reasonable, since $T_{\rm min} \propto \Delta^2$ and $H\propto T_g^{1/2}$,if the disk is vertically in a hydrostatic equilibrium.
In Fig. \ref{fig: f5}, we compare density PDF in the two models
with different spatial resolution.
PDFs in the models are qualitatively similar, in a sense that
the PDF has a LN part for high density gas, 
although the PDF in the model with a 10 pc resolution 
is not perfectly fitted by a LN function.
 This comparison implies that the  
spatial resolution should be at least
5 pc in order to discuss PDF of the ISM in galactic disks.


Figure \ref{fig: phase} shows a phase-diagram of model D.
Two dominant phases in volume, i.e. warm gas $T_g \sim 8000$ K and cold gas ($T_g \sim 30-1000$K) exist. Temperature of the gas denser 
than $\rho \simeq 10 M_\odot$ pc$^{-3}$ 
is limitted by $T_{\rm min}$. 
As suggested by Fig. \ref{fig: phase},
frequency distribution of temperature is not represented by
a single smooth function, which is a notable difference to density PDF.
High density gas ($\rho > 10^{-2} M_\odot$ pc$^{-3}$) is not
isothermal, suggesting that  isothermality is not a necessary 
condition for the LN density PDF in global models of the ISM (see also 
\S 4.1).

As qualitatively seen in Fig. \ref{fig: f1}, the dispersion of LN-PDF
is larger in more massive system. This is more quantitatively 
seen in Fig. \ref{fig: fmass}, mass-weighted PDFs in 4 models with 
different $\rho_i$. In the fit, we use the same $\sigma$ obtained from
the volume-weighted PDF, and 
the mass-weighted characteristic density $\rho_{0,M}$ is calculated using
\bea
\rho_{0,M} = \rho_{0}e^{\sigma^2}.
\label{eqn: eq6}
\eea
Equation (\ref{eqn: eq6}) is always true if 
the PDF is log-normal. 
The dispersion $\sigma$ is 2.36 in model A ($\rho_i = 5 M_\odot$ pc$^{-3}$)
and 3.01 in model D ($\rho_i = 50 M_\odot$ pc$^{-3}$) (Table 1).
The dependence of $\sigma$ on the initial gas density (or total gas mass)
is a natural consequence of LN-PDF (see \S 3.1).
From comparison between the volume-weighted and 
mass-weighted PDF, it is clear that 
even if the smooth non-LN regime occupy 
the most part of the volume of the galactic disk-halo region (i.e. $\alpha < 1$),
the mass of ISM is dominated by the LN-PDF regime.

Another important point in the numerical results is that
the characteristic density of LN-PDF, $\rho_0$,
does not significantly change among the
models, despite the wide (almost 8 orders of magnitude) 
density range in the quasi-steady state.
In fact, as shown in Table 1, $\rho_0$ is in a range of $1.7<\rho_0< 2.7 $ cm$^{-3}$
among the models with $\rho_i = 5-50 M_\odot$ pc$^{-3}$.
Physical origin of this feature is discussed in \S 3.1.

In summary, using our three-dimensional high resolution 
hydrodynamic simulations, we find that the ISM in a galactic disk is 
inhomogeneous on a local scale, 
and it is consisted of many filaments, clumps, and
low-density voids, which are in a quasi-steady state on a global scale.
Statistical structure of the density field in the galactic disks
is well described by
a single LN-PDF over 6 decades in a high density part (i.e. $\rho \gtrsim 0.01 M_\odot$ pc$^{-3}$). Most part of the mass is in the regime of LN-PDF.
The dispersion of the LN-PDF is larger for more massive system.

%
\section{PROBABILITY DISTRIBUTION FUNCTION AND THE STAR FORMATION RATE}
%
\subsection{Basic Properties of the Log-Normal PDF}
In this section, based on properties of LN-PDF, we discuss
how we can understand the numerical results in \S 2.
Suppose that the density PDF, $f(\rho)$ in the galactic disk
is described by a single log-normal function:
\bea
f(\rho)d\rho = \frac{1}{\sqrt{2\pi}\sigma} 
\exp{\left[
-\frac{
\ln (\rho/\arho)^2
}{
2\sigma^2
}
\right] } d \ln \rho,
\eea
where $\arho$ is the characteristic density and $\sigma$ is the 
dispersion. 
The volume average density $\langle \rho\rangle_V$ for the gas described by a
the LN-PDF is then
\begin{eqnarray}
\langle\rho\rangle_{V} &=& \frac{1}{\sqrt{2\pi}\sigma} 
\int^\infty_{-\infty}
\rho
\exp{\left[
-\frac{
\ln (\rho/\arho)^2
}{2\sigma^2}
\right] } d \ln \rho,  \\
&=& \rho_0 \, e^{{\sigma^2}/{2}}. \label{eq: msigma}
\end{eqnarray}
The mass-average density is 
\bea
\langle\rho\rangle_M  &\equiv& \frac{\int \rho^2 dV}{M_t} = \rho_{0,M} e^{\sigma^2} = \rho_{0} e^{2\sigma^2} 
\label{eq: massavr}
\eea
where $M_t = \int \rho dV$ is the total mass
in the log-normal regime.
The dispersion $\sigma$ of the LN-PDF is therefore
\bea
\sigma^2 = \frac{2}{3} \ln \left( \frac{\langle\rho\rangle_M}{\langle\rho\rangle_V} \right).
\eea
Equivalently, using $\rho_0$, 
\bea
\sigma^2 &=& 2 \ln \left(\frac{\langle\rho\rangle_V}{\rho_0} \right) \label{eqn: eq17a}\\ 
 &=& \frac{1}{2} \ln \left(\frac{\langle\rho\rangle_M}{\rho_0}\right).
\label{eqn: eq17b}
\eea
Suppose the characteristic density $\rho_0$ is nearly constant
as suggested by the numerical results,
Equations (\ref{eqn: eq17a}) or (\ref{eqn: eq17b}) tells that the dispersion $\sigma$ is larger for more massive systems, which is also consistent with the numerical results.
For a stable, uniform system, i.e. $\langle{\rho}\rangle_V = \arho$, $\sigma$ should be zero.
In another extreme case, namely $\langle{\rho}\rangle_V \rightarrow \infty$, $\sigma \rightarrow \infty$, but this is not the case, because the system itself is no longer dynamically stable.
Therefore, $\sigma$ should take a number in an appropriate range in a globally stable, inhomogeneous systems. Galactic disks are typical examples of such systems.

In our numerical results, the ISM with low density part (typically less than
$10^{-3} M_\odot$ pc$^{-3}$) is not fitted by the LN-PDF.
If density field of the ISM in a fraction, $\alpha$, of the  arbitrarily volume is characterized by the LN-PDF, the volume average density 
in the volume is $\bar{\rho} = \alpha \langle\rho\rangle_V$. 
In this case, eq. (\ref{eqn: eq17a}) is modified to
\bea
\sigma_p^2 = 2 \ln \left(\frac{\bar{\rho}}{\alpha \rho_0}\right).
\label{eqn: eq17c}
\eea
As shown in Table 1, $\sigma$ in the numerical results 
well agrees with $\sigma_p$ in each model.
In numerical simulations, it is easy to know
the volume in the LN-regime (i.e. $\alpha \times$ volume of 
the computational box), therefore calculating $\sigma_p$ is straightforward
using eq. (\ref{eqn: eq17c}).
However, if one wants to evaluate $\sigma$ from observations of galaxies,
it is more practical to use the mass-average density, eq. (\ref{eqn: eq17b}),
because it is expected that mass of the galactic ISM is dominated by
 the LN-regime in mass (see Fig. \ref{fig: fmass}).

\begin{figure}[t]
\centering
\includegraphics[width = 8cm]{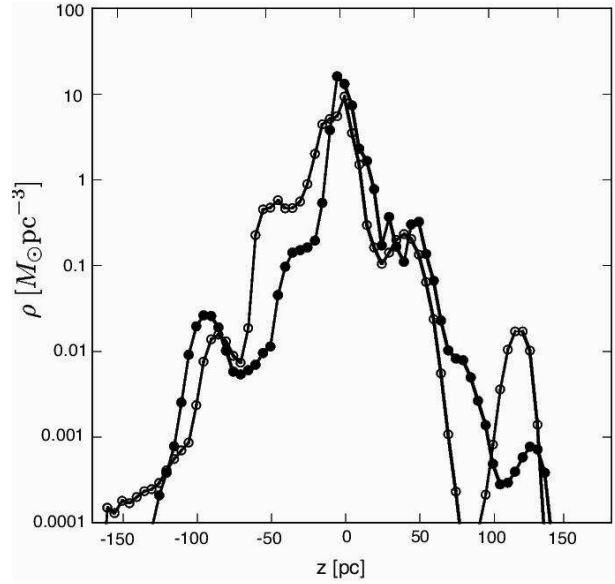}
\caption{Vertical density structure of model D.  Two lines represent
density averaged for x-direction (black dots) and y-direction (open circles).}
\label{fig: fig_vertical}
\end{figure}

\begin{figure}[t]
\centering
\includegraphics[width = 8cm]{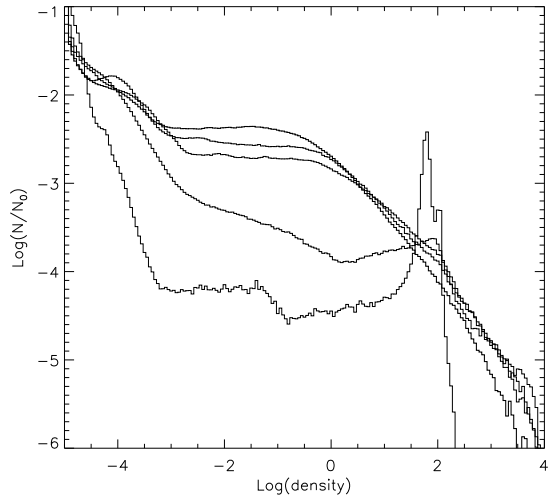}
\caption{Time evolution of density PDF for the whole
computational volume for model D. 
The lines represent the PDF at $t=3, 10, 16, 21$, and 43 Myr (thick solid line). The vertical axis is the number fraction of 
grid cells for the density. The unit of density is $M_\odot$ pc$^{-3}$.}
\label{fig: f2}
\end{figure}

\begin{figure}[t]
\centering
\includegraphics[width = 8cm]{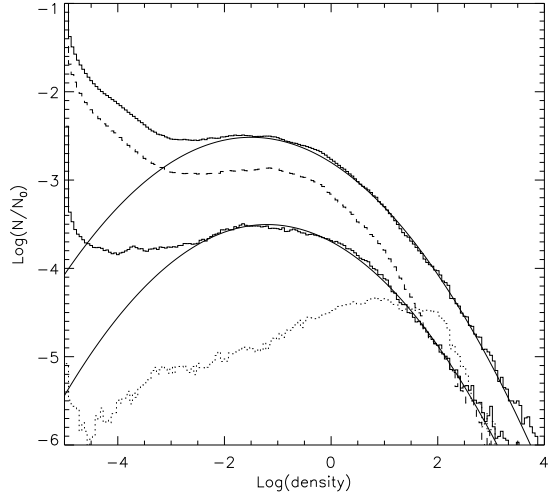}
\caption{Density PDF for four sub-regions in model D. 
Solid line is  a thin disk ($|z| \leq 10$ pc),
dotted line is a inner disk 
($10 \leq z \leq 20 $ pc and $|x,y| \leq 0.38$ kpc),
dashed line is a halo ($30 \leq z \leq $ 160 pc),
where $z=0 $ is the galactic plane. The two solid curves are
a log-normal function with $\sigma_{10} = 1.273$, $\log(\arho) = -1.2$
and $\alpha =$ 0.02 (disk), and
$\sigma_{10} = 1.308$, $\log(\arho) = -1.5$
and $\alpha =$ 0.2 (the whole region).  }
\label{fig: f3a}
\end{figure}

\begin{figure}[t]
\centering
\includegraphics[width = 8cm]{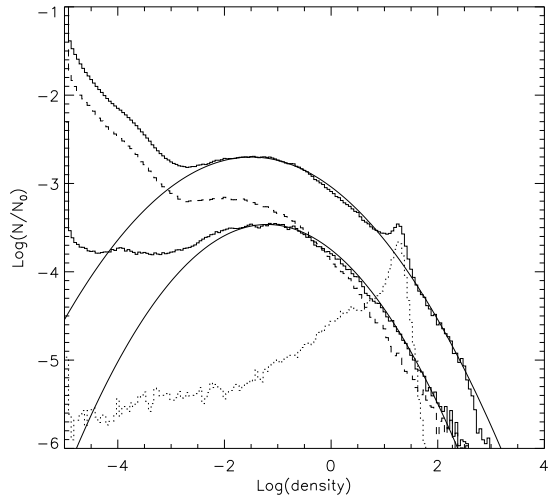}
\caption{Same as Fig. \ref{fig: f3a}, but for model B.}
\label{fig: f3b}
\end{figure}

\begin{figure}[t]
\centering
\includegraphics[width = 8cm]{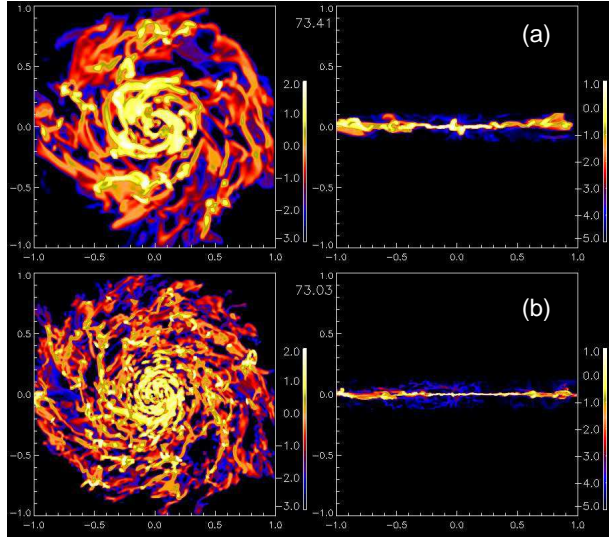}
\caption{Density structures of model C with two deferent resolutions.
A temperature floor depending on the resolution: 
(a) $\Delta = 10$ pc, and (b) $\Delta = 5$ pc.
$T_{\rm min} = 35$ K ($\rho_i/10 M_\odot$ pc$^{-3}$) ($\Delta$/5 pc)$^2$ is
assumed to ensure that the Jeans unstable is resolved. 
}
\label{fig: f4}
\end{figure}

\begin{figure}[t]
\centering
\includegraphics[width = 8cm]{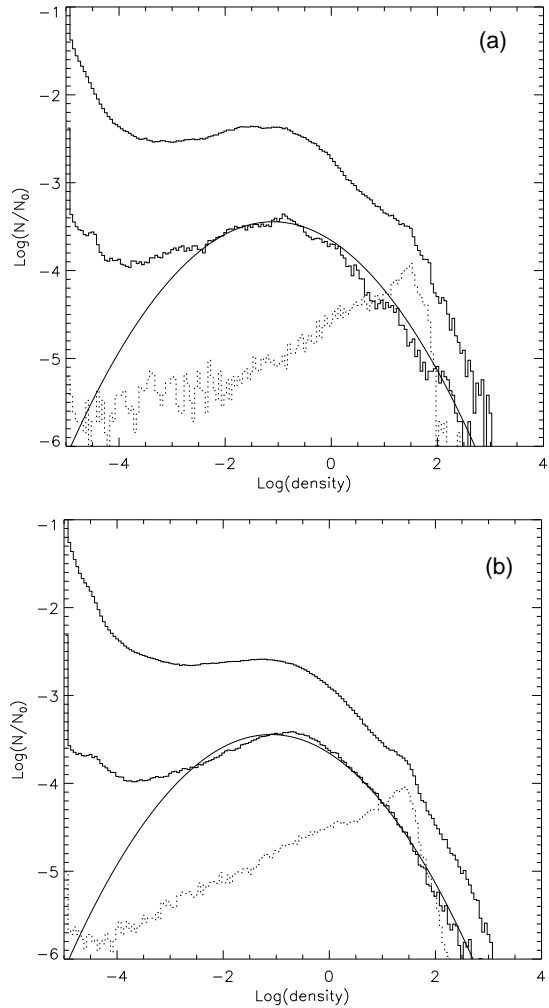}
\caption{Density PDF for model C with two deferent resolutions.
 (a) $\Delta = 10$ pc, (b) $\Delta = 5$ pc. 
Three lines are PDF for the whole computational box, disk, and the inner
disk (dotted line).}
\label{fig: f5}
\end{figure}

\begin{figure}[t]
\centering
\includegraphics[width = 8cm]{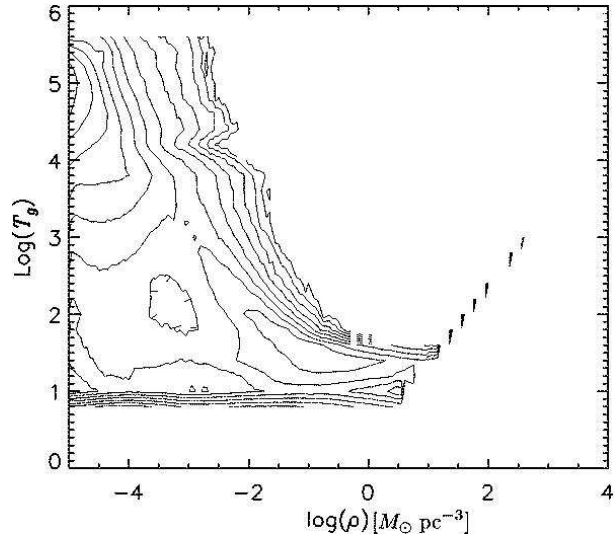}
\caption{Phase diagram of model D at $t= 43$ Myr. Contours 
represent log-scaled volumes.}
\label{fig: phase}
\end{figure}

\begin{figure}[t]
\centering
\includegraphics[width = 9cm]{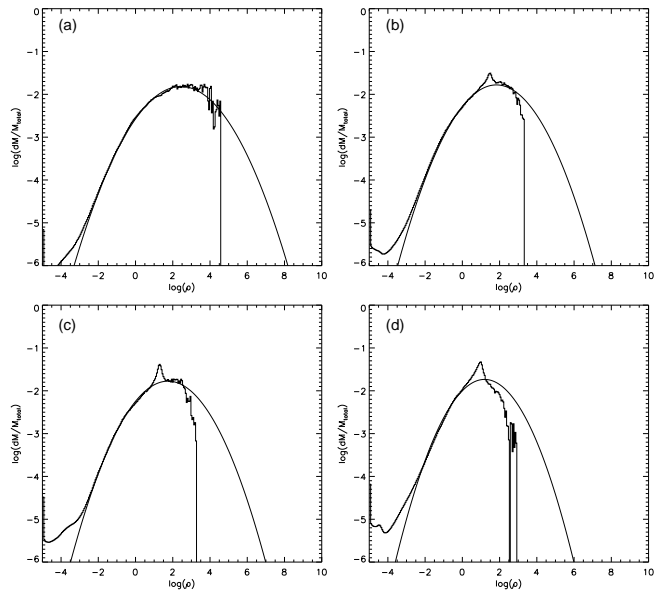}
\caption{Mass-weighted PDF for (a) model D, (b) model C, (c) model B, and (d) model A.  Unit of density is $M_\odot$ pc$^{-3}$.
 In the log-normal fit, $\rho_{0,M} = \rho_{0}e^{\sigma^2}$ (eq. (\ref{eqn: eq6})) and $\sigma$ for the volume-weighted PDF (table 1) are used. 
The cut-off at high density in each plot is caused by the resolution limit.
}
\label{fig: fmass}
\end{figure}

\begin{figure}[t]
\centering
\includegraphics[width = 8cm]{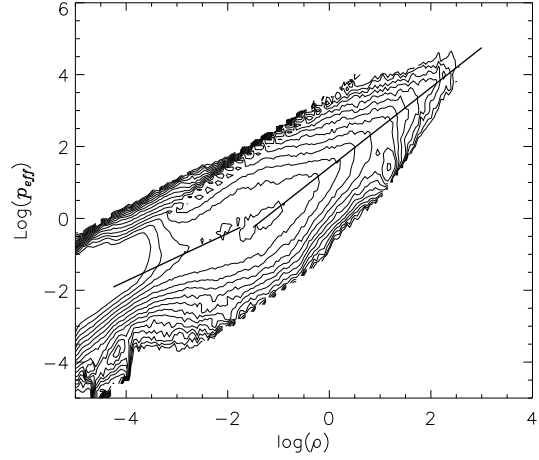}
\caption{Effective pressure-density diagram of model E. Two power-laws are $p_{\rm eff} \propto \rho^{2/3}$ and $p_{\rm eff} \propto \rho^{4/3}$. $p_{\rm eff} \equiv p_{\rm th} + \rho v_t^2$, where $p_{\rm th}$ is the thermal pressure and $v_t$ is turbulent velocity of the medium.
The contours represent log-scaled volumes for given density and $p_{\rm eff}$.}
\label{fig: rhop}
\end{figure}

Interestingly, even if the density contrast is extremely large (e.g. $10^6-10^8$),
the numerical results show that the characteristic density $\rho_0$ 
is not sensitive for the total gas mass in a kpc galactic disk
($\rho_0 = 1.7-2.7$ cm$^{-3}$).
If this is the case, what determines $\rho_0$? 
In Fig. \ref{fig: rhop}, we plot effective pressure, $p_{\rm eff}\equiv
p_{\rm th} + \rho v_t^2$ as a function of density, where $p_{\rm th}$ is 
the thermal pressure, and $v_t$ is turbulent velocity dispersion,
which is obtained by averaging the velocity field in each sub-region 
with a (10 pc)$^3$ volume.
If the turbulent motion in a volume with a size $L$ 
is originated in selfgravity of the gas and galactic rotation\citep[][]{wad02b}, $p_{\rm eff} \sim \rho G M_g/L \propto \rho^{4/3}$, provided that the mass in 
the volume $M_g$ is conserved (i.e. $\rho  \propto L^{-3}$)\footnote{$L$ is not 
a size of molecular clouds, but size of `eddies' turbulent motion in
the inhomogeneous ISM. \citep[See Fig. 2 in][for example]{wad02b}.
Therefore, the Larson's law, i.e. $\rho \propto $ (size of molecular clouds)$^{-1}$, is not a relevant scaling relation here.}.
In fact, Fig. \ref{fig: rhop} shows that a majority of the gas in the dense part 
($\rho \gtrsim 10^{-1} M_\odot$ pc$^{-3}$) follows $p_{\rm eff} \propto \rho^{4/3}$.
 On the other hand, for the 
low density regime, $p_{\rm eff} \propto \rho^{2/3}$. 
In the present model, a dominant heating source in low density gases
is shock heating. Shocks are ubiquitously generated by 
turbulent motion, whose velocity is $\sim$ several 10 km s$^{-1}$. 
The kinetic energy is thermalized at shocks, then
temperature of low density gas goes to $T_g \sim 10^{4-5}$ K.
In supersonic, compressible turbulence, 
its energy spectrum is expected to be $E(k) \propto k^{-2}$,
where total kinetic energy $E_t = \int E(k) dk$.
Therefore $p_{\rm eff} \propto \rho v^2 \propto \rho^{2/3}$, 
because $v^2 \propto k^{-1} \propto \rho^{-1/3}$.
Then the `effective' sound velocity $c_{\rm eff}^2 \equiv dp_{\rm eff}/d\rho \propto
\rho^{-1/3}$ and $c_{\rm eff}^2 \propto \rho^{1/3}$ for the low and high density
regions, respectively.
This means that the effective sound velocity $c_{\rm eff}$ has a minimum
at the transition density ($\rho_t$) between the two regimes. In other words, since $c_{\rm eff}^2 = c_s^2 + v_t^2$,
there is a characteristic density below/above which
 thermal pressure/turbulent pressure dominates the total pressure,
i.e. $c_s^2 \sim v_t^2$. Thus,
\bea
\frac{k T_g}{\mu} \sim \frac{GM_g}{L} \sim G\rho_t L^2, 
\eea
where $\mu$ is average mass per particle, and $L$ is size of the largest
eddy of gravity-driven turbulence, which is roughly the scale height of the disk.
As seen in Fig. \ref{fig: phase}, gas temperature around $\rho_0 = 10^{-1.5} M_\odot$ pc$^{-3}$ is $T_g \sim 80$ K. 
The scale height of the dense disk is about 10 pc (Fig. \ref{fig: fig_vertical})
 in the present model,
therefore
\bea
\rho_t \simeq 1 \;\;\; {\rm cm}^{-3} \left(\frac{T_g}{80 \; {\rm K}}\right)
\left(\frac{L}{10 \; {\rm pc}}\right)^{-2}.
\label{eq: rhot}
\eea
This transition density is close to the characteristic density
in the numerical results, i.e. $\rho_0 = 10^{-1.4} - 10^{-1.6} M_\odot$ pc$^{-3}$($ \approx $ 2.7 - 1.7 cm$^{-3}$).
In a high-density region ($\rho > \rho_t \approx \rho_0$),
stochastic nature of the system, which is the origin of
the log-normality, is caused in the gravity-driven turbulence,
and for low density regions its density field is randomized 
by thermal motion in hot gases.
For the gases around $\rho_0$, the random motion is relatively static,
therefore probability of the density change is small.
As a result, the density PDF takes the maximum around $\rho_0$.
An analogy of this phenomenon is a snow or sand drift in a turbulent air.
The material in a turbulent flow is stagnated in a
relatively ``static'' region.

As seen in Fig. \ref{fig: fig_vertical}, the present gas disk has
a much smaller scale height than the ISM in real galaxies, 
which is about 100 pc. If turbulent motion in galactic disks
is mainly caused by gravitational and thermal instabilities in 
a rotating disk \citep{wad02b}, we can estimate $\rho_t$ by
eq. (\ref{eq: rhot}) in galactic disks, but gas temperature of
the dominant phase in volume is $T_g \sim 10^4$ K, 
as suggested by two-dimensional models \citep{wad01b}.
Therefore, we expect that $\rho_t \sim 1$ cm$^{-3}$ is also the case
in real galaxies. However, if the turbulent motion in ISM is 
caused by different mechanisms, such as supernova explosions, 
magneto-rotational instabilities, etc.,  dependence of $p_{\rm eff}$ 
on density could be different from $p_{\rm eff} \propto \rho^{4/3}$, 
and as a result $\rho_t \sim$ const. $\sim O(1)$ cm$^{-3}$ 
could not be always true.
Determining $\rho_0$ in a much larger disk than the present model,
taking into account various physics is an important problem for 3-D simulations with a 
large dynamic range in the near future.
In \S 3.3, we compare our results with SFR in spiral and 
starburst galaxies, in which  $\rho_0$ is treated as one of free parameters
(see Fig. \ref{fig: sfr-rho0} and related discussion).

Suppose the volume average density is $\langle\rho\rangle_V = 3 M_\odot $ pc$^{-3}$, 
we can estimate the dispersion $\sigma$ is $\sigma \simeq 3$ for $\rho_0 = 1$ cm$^{-3}$ 
from eq. (\ref{eqn: eq17a}).
For a less massive system, e.g. $\langle\rho\rangle_V = 0.3 M_\odot $ pc$^{-3}$, 
$\sigma \simeq 2.1$. 
Therefore we can expect a larger star formation rate for
more massive system, since a fraction of high density gas is larger.
Using this dependence of $\sigma$ on the average gas density,
we evaluate global star formation rate in the next section.


\subsection{Global Star Formation Rate}
Based on the numerical results in \S 2 and the properties of LN-PDF
described in \S 3.1, we here propose a simple theoretical model of the
star formation on a global scale. We assume that the multi-phase,
inhomogeneous ISM in a galactic disk, 
can be represented by a LN-PDF.
\citet{elm02} discussed  a fraction of 
high density gas and global star formation rate assuming 
a LN-PDF.
Here we follow his argument more specifically. 

If star formation is led by gravitational collapse of high-density clumps
with density $\rho_c$, 
the star formation rate per unit volume, $\dot{\rho_\star}$, can be written as
\bea
\dot{\rho_\star} = \epsilon_c (G\rho_c)^{1/2} f_c \langle\rho\rangle_V
\label{eqn: sfr}
\eea
where $f_c(\rho_c, \sigma)$ is a mass fraction of the gas whose density is 
higher than a critical density for star formation ($\rho > \rho_c$),
and $\epsilon_c$ is the efficiency of the star formation.

\begin{figure}[p]
\centering
\includegraphics[width = 8cm]{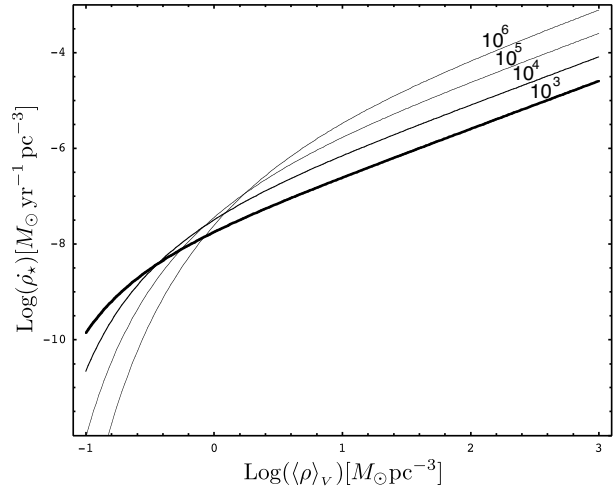}
\caption{Star formation rate ($\dot{\rho_\star}$)
as a function of volume average density ($\bar{\rho}$)
for 0.01.
$\rho_0 = 1$ cm$^{-3}$ is assumed.
Four curves are plotted for $\delta_c = 10^3$ (thick solid line), $10^4, 10^5$ and $10^6$ (from thick to
thin lines) are plotted.}
\label{fig: sfr-vol}
\end{figure}

\begin{figure}[p]
\centering
\includegraphics[width = 8cm]{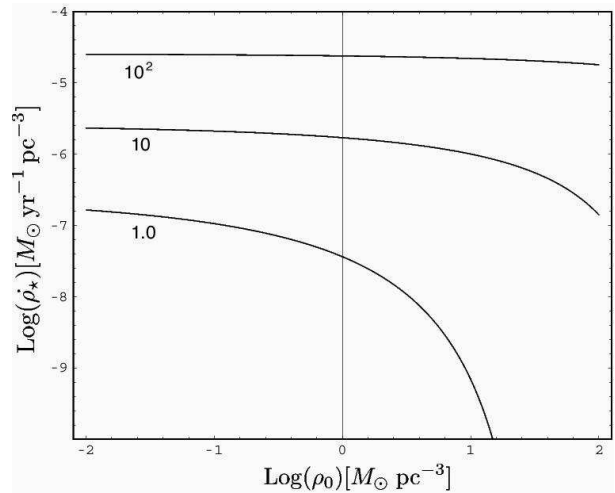}
\caption{Star formation rate ($\dot{\rho_\star}$) as a function of the characteristic density of the LN-PDF ($\rho_0$) for different average density $\langle\rho\rangle_V = 10^2, 10$, and $1 M_\odot$ pc$^{-3}$.
 $\delta_c = 10^4$ is assumed.}
\label{fig: sfr-rho0}
\end{figure}

Suppose the ISM model found in \S 2, whose density field is
characterized by LN-PDF,  is applicable to the ISM in galactic 
disk, $f_c$ is

\bea
f_c (\delta_c, \sigma) &=& \frac{ \int^\infty_{\ln \delta_c} \delta
\exp\left[ - \frac{(\ln \delta)^2}{2\sigma^2}\right] d(\ln \delta)
}
{
\int^\infty_{-\infty} \delta \exp\left[ - \frac{(\ln \delta)^2}{2\sigma^2}\right] d(\ln \delta)
}, \\
&=& \frac{1}{2}(1 - {\rm Erf}[z(\delta_c,\sigma)])\;,
\label{eq: fc}
\eea
where $\delta \equiv \rho/\arho$, $\delta_c \equiv \rho_c/\arho$, and
\bea
z(\delta_c,\sigma) \equiv \frac{\ln \delta_c - \sigma^2}{\sqrt{2} \sigma}.
\label{eq: zdelta}
\eea
The fraction of dense gas, $f_c$ is a monotonic function of $\delta_c$ and $\sigma$, and it decreases
rapidly for decreasing $\sigma$.
Suppose $\delta_c = 10^5$, $f_c \sim 10^{-2}$ for $\sigma = 3.0$, and 
$f_c \sim 10^{-6}$ for $\sigma = 2.0$\footnote{
In \citet{elm02}, the dispersion of the LN-PDF,
 $\sigma = 2.4$ is assumed,
which is taken from 2-D hydrodynamic simulations of the multi-phase 
ISM in WN01.}.
The star formation rate (eq. (\ref{eqn: sfr})) per unit volume then can be rewrite as a
function of $\epsilon_c, \delta_c$, and $\sigma$:
\bea
\dot{\rho}_\star(\epsilon_c, \delta_c, \sigma) &=&  \epsilon_c (G\delta_c)^{1/2} f_c \arho^{3/2} e^{\sigma/2}
\label{eqn: eq-24}
\eea
Using equations (\ref{eqn: eq17a}), (\ref{eq: fc}) and (\ref{eq: zdelta}),
we can write SFR as a function of 
volume-average density $\langle \rho\rangle_V$,
\bea
\dot{\rho}_\star
\left[
\epsilon_c,
 \left(\frac{\langle\rho\rangle_V}{1M_\odot {\rm pc}^{-3}}\right),
 \left(\frac{\rho_0}{1\,{\rm cm}^{-3}}\right),
 \left(\frac{\rho_c}{10^5\,{\rm cm}^{-3}}\right)
\right] 
= 3.6 \times 10^{-7} \epsilon_c \; M_\odot {\rm yr}^{-1} 
{\rm pc}^{-3} \label{eqn: eq-25}\\  \nonumber
\times \left[1-{\rm Erf}\left(\frac{\ln(\rho_{c}\rho_0/\langle\rho\rangle^2_V)}
{2 [\ln(\langle\rho\rangle_V/\rho_0)]^{1/2}} \right)\right] \rho_{c}^{1/2} \langle\rho\rangle_V.
\label{eq: modelsfr}
\eea
We plot eq.(\ref{eqn: eq-25}) as a function of the volume average density in Fig. 
\ref{fig: sfr-vol}. Four curves are plotted for
 $\rho_c = 10^3, 10^4, 10^5$ and $10^6$ cm$^{-3}$.
We can learn several features from this plot. SFR increases rapidly for
increasing the average density, especially for lower average 
density ($ < 100 \, M_\odot$ 
pc$^{-3}$) and higher critical density.
This behavior is naturally expected for star formation in the ISM
described by LN-PDF, because the dispersion $\sigma$ changes
logarithmically for the average gas density (eq. (\ref{eqn: eq17a})).
 For large density, it approaches to SFR $\propto \langle\rho\rangle_V$. 
SFR does not strongly depend on
$\rho_c$ around $\langle{\rho}\rangle_V \sim 1 M_\odot$ pc$^{-3}$, and SFR is
larger for larger $\rho_c$ beyond the density. This is because the free fall
time is proportional to $\rho_c^{-1/2}$, and this compensates decreasing 
the fraction of high density gas, $f_c$.

Fig. \ref{fig: sfr-rho0} shows how SFR changes as a function
of $\rho_0$. SFR does not strongly depend on $\rho_0$ around $\rho_0 \lesssim 0.1
M_\odot $ pc$^{-3}$ ($\simeq 7$ cm$^{-3}$).
 See also Fig. \ref{fig: sfr-f14} in terms of 
comparison with observed SFR.

\subsection{Comparison with Observations}
The theoretical SFR based on the LN-PDF in the previous section has a 
couple of free parameters.
Comparison between the model and observations is useful to narrow down
the parameter ranges.

We should note, however that the gas disk presented in \S 2 is 
geometrically thiner ($\simeq 10$ pc) than the real galactic disk ($\simeq 100 $ pc).  In this sense, although the present models are based on 
full 3-D simulations, they are not necessarily adequate
for modeling typical spiral galaxies\footnote{One can 
refer to a critical study for modeling galactic ISM using 
a 2-D approximation \citep{sanc01}}.
 There are couple of reasons why the present model has small scale height.
Since the disks here are relatively small ($r \simeq 1$ kpc), 
the disks tend to be thin due to the deep gravitational potential.
Self-gravity of the gas also contributes to make the disks thin.
Radiative energy loss in the high density gas in the central region 
cancels the energy feedback from the supernovae (see \S 4.5).
Therefore, a simple way to make the model gas disks thicker to fit real
galactic disks is simulating a larger disk (e.g. $r\sim 10$ kpc)
with an appropriate galactic potential and supernova feedback,
solving the same basic equations\footnote{
This is practically difficult, if a pc-scale spatial resolution 
is required. A recent study by  \citet{task06} is
almost the only work that can be directly compared with
real galactic disks. Unfortunately, \citet{task06} do not 
discuss on the density PDF, but they take a different approach
to study SFR by generating ``star'' particles (see \S 1).
They claimed that a Schmidt-law type SFR is reproduced in their
simulations. This is consistent with
our analysis, provided that the density PDF is log-normal like, and
that local star formation takes place above a critical density.}.
Besides supernova explosions, 
there are possible physical processes to puff up the disks,
such as nonlinear development of 
magneto-rotational instability, 
heating due to stellar wind and strong radiation field from 
star forming regions. 
These effects should be taken into account in 3-D simulations
with high resolutions and large dynamic ranges in the near future, and 
it is an important subject whether the LN-PDF is reproduced in such more realistic
situations. Another interesting issue in terms of PDF is effect of 
a galactic spiral potential. 
However, here we try to make a comparison with observations, 
assuming that the log-normal nature in the
density field found in our simulations is also the case in real galactic
disks. This would not be unreasonable, because 
log-normality is independent of the geometry of the system.
In fact a LN-PDF is also found in a 3-D torus model for AGN \citep{wad05}.
A LN-PDF is naturally expected, if the additional physical processes causes non-linear,
random, and independent events that change the density field.

\begin{figure}[p]
\centering
\includegraphics[width = 8cm]{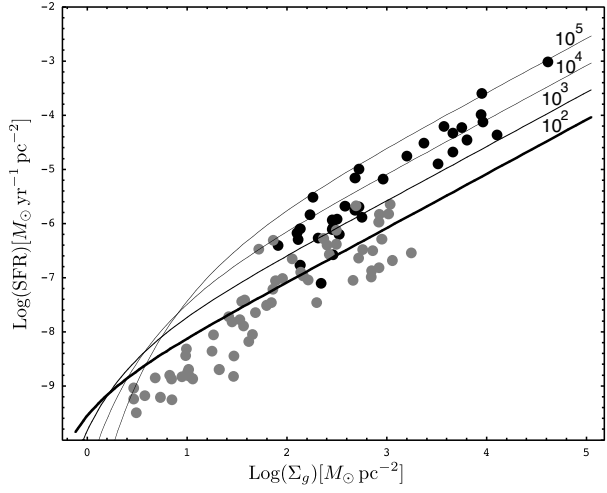}
\caption{Comparison between our models and observed surface star formation rate in terms of surface gas density. Black dots are starburst galaxies in 
Kennicutt (1998) and gray dots are normal galaxies in \citet{kom05}.
Four curves are SFR for $\delta_c = 10^2$ (thick solid line), $10^3, 10^4$ and $10^5$.
$\alpha = 0.1$, $\rho_0 = 1.0$, and $\epsilon_c = 0.01$ are assumed.
}
\label{fig: sfr-f13}
\end{figure}

\begin{figure}[p]
\centering

\includegraphics[width = 8cm]{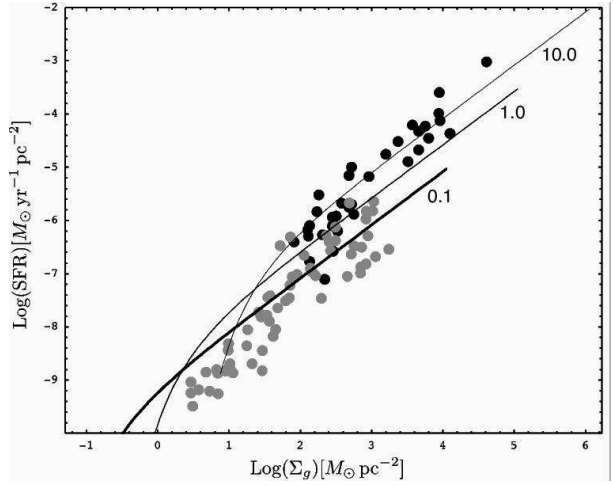}
\caption{Same as Fig. \ref{fig: sfr-f13}, but for dependence of
characteristic density of LN-PDF, $\rho_0 = 0.1$ (thick solid line), 1 and $10 M_\odot$ pc$^{-3}$.
$\alpha = 0.1$, $\delta_c = 10^3$, and $\epsilon_c = 0.01$ are assumed.}
\label{fig: sfr-f14}
\end{figure}

\begin{figure}[p]
\centering
\includegraphics[width = 8cm]{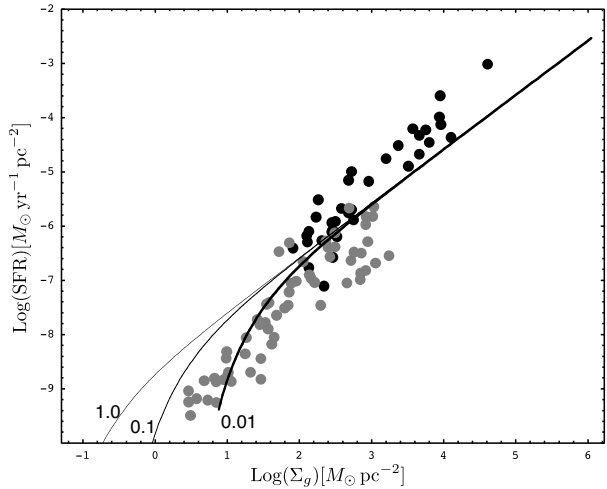}
\caption{Same as Fig. \ref{fig: sfr-f13}, but for dependence of
fraction of LN part, $\alpha= 0.01$ (thick solid line), 0.1, and 1.0.
$\rho_0=1.0$, $\delta_c = 10^3$, and $\epsilon_c = 0.01$ are assumed.}
\label{fig: sfr-f15}
\end{figure}

Figure \ref{fig: sfr-f13} shows 
surface star formation rate $\dot{\Sigma}_\star$ ($M_\odot$ yr$^{-1}$ pc$^{-2}$) 
in normal and starburst galaxies \citep{kom05, ken98}.
The scale height of the ISM is assumed to be 100 pc.
Four curves represent model SFR with different critical density.
It is clear that smaller $\delta_c$ is preferable to explain
the observations, especially for low average density. For example, SFR with $\delta_c = 10^5$ is too steep.
Similarly, Fig. \ref{fig: sfr-f14} shows dependence of SFR on the characteristic
density $\rho_0$. As mentioned above, SFR is not very sensitive for
changing $\rho_0$, but $\rho_0 \lesssim 1.0$ cm$^{-3}$ is better to explain the
observations.  From Fig. \ref{fig: sfr-f15}, which is
how SFR depends on a fraction of LN part in volume, $\alpha$.
SFR is sensitive for $\alpha$ especially for low density media.
The plot implies that $\alpha = 0.01$ is too steep to
fit the observations, especially for normal galaxies.

\begin{figure}[p]
\centering
\includegraphics[width = 8cm]{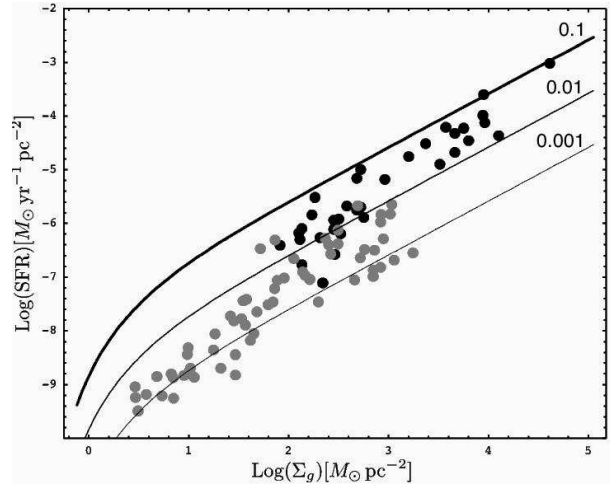}
\caption{Same as Fig. \ref{fig: sfr-f13}, but for
the best-fit model ($\rho_0 = 1.0, \alpha = 0.1,$ and $\delta_c=10^3$).
Three curves are for $\epsilon_c = 0.1$ (thick solid line), 0.01 and 0.001.
}
\label{fig: sfr}
\end{figure}

After exploring these parameters, we find that 
Fig. \ref{fig: sfr} is the best-fit model
with $\rho_0 = 1$ cm$^{-3}$,  $\alpha = 0.1$ (eq. (\ref{eqn: eq17b}))
and $\delta_c = 10^3$.
Four curves corresponds to SFR with
efficiency, $\epsilon_c = 0.1$, 0.01, and 0.001.
The model slope approaches to SFR $\propto \Sigma_g$ for large density,
 which is shallower than the Kennicutt law (i.e.  SFR $\propto \Sigma_g^{1.41}$).
Observed SFRs in most galaxies are distributed between the model curves with
$\epsilon_c \simeq 0.1$ and $0.001$. The starburst galaxies are distributed 
in $\epsilon_c = 0.1-0.01$, on the other hand, the normal galaxies \citep{kom05}
show systematically smaller SFR, which is consistent with
smaller efficiency (i.e $\epsilon_c = 0.01-0.001$)
 than those in starburst galaxies.
This suggests that the large SFR in starburst galaxies is achieved
 by both high average gas density and large (several \%) star formation efficiency
in dense clouds\footnote{An alternative explanation can be possible: the typical 
scale height of the star forming regions is different by a factor of 10-100 in 
normal and starburst galaxies for the same efficiency.}.

\begin{figure}[p]
\centering
\includegraphics[width = 8cm]{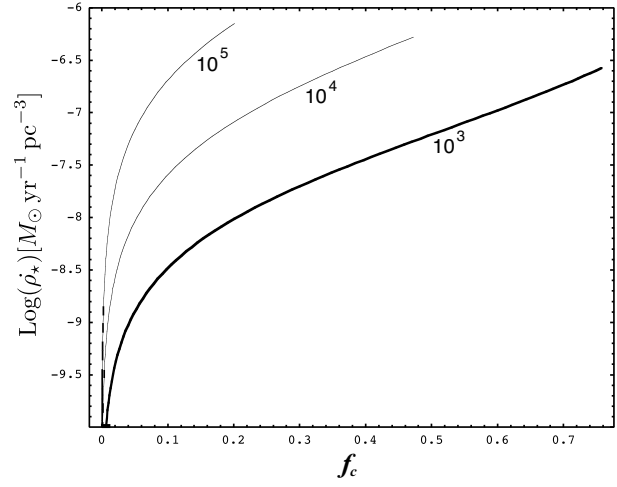}
\caption{SFR as a function of a fraction of high density gas $f_c$. Three curves
are plotted for $\delta_c = 10^3$ (thick line), $10^4$, and $10^5$.
$\epsilon_c = 0.1$ and $\alpha = 0.1$ are assumed.}
\label{fig: sfr_fc}
\end{figure}

\citet{gao04} found that there is a clear positive correlation between 
HCN and CO luminosity in 65 infrared luminous galaxies and normal spiral galaxies.
They also show that luminous and ultra-luminous infrared galaxies tend
to show more HCN luminous, suggesting a larger fraction of dense
molecular gas in active star forming galaxies.
Figure \ref{fig: sfr_fc} is SFR as a function of $f_c$ using the model in
the previous section. Interestingly, 
a qualitative trend of this plot is quite similar to 
that of Fig. 4 in \citet{gao04},
which is SFR as a function of $L_{\rm HCN}/L_{\rm CO} (\propto$ dense gas
fraction).  Both observations and our model show that 
SFR increases very rapidly for increasing $f_c$, especially
when $f_c$ or $L_{\rm HCN}/L_{\rm CO}$ is small ($< 0.1$)
and then it increases with a power-law.





%
\section{DISCUSSION}
%
\subsection{What Determines the Dispersion of LN-PDF?}

Elmegreen (2002) first pointed out that if the density 
PDF in the ISM is described by a log-normal function, 
the star formation rate should be a function of the critical density for local
star formation and the dispersion of the LN-PDF. The critical 
density for local star formation should be determined not
only by gravitational and thermal instabilities of the ISM, but 
also by magneto-hydrodynamical, chemical and radiative processes on a pc/sub-pc scale.
It is beyond a scope of the present paper that
how the critical density is determined. 
The other important parameter, $\sigma$ in LN-PDF, should be
related to physics on a global scale.

Some authors claimed that the LN-PDF is a characteristic 
feature in an isothermal, turbulent flow, and its dispersion
is determined by the rms Mach number \citep{vazq94, pado97, nord98, scal98}.
This argument might be correct for the ISM on a local scale, for example
an internal structure of a giant molecular cloud, which is nearly isothermal
and a single phase. 
\citet{krum05} derived an analytic prediction for the star 
formation rate assuming that star formation occurs in virialized 
molecular clouds that are supersonically turbulent and that density 
distribution within a cloud is log-normal.
In this sense, their study is similar to Elmegreen (2002) and the present work.
However, they assumed that the dispersion of the LN-PDF is a 
function of ``one-dimensional Mach number'', ${\cal M}$,  of the turbulent motion, 
i.e. $e^{\sigma^2} \approx 1+3{\cal M}^2/4$,
which is suggested by numerical experiments of isothermal turbulence
 (Padoan \& Nordlund 2002). 
A similar empirical relation between the density contrast and magneto-sonic Mach number
is suggested by Ostriker, Stone \& Gammie (2001).
Yet the physical reason why the dispersion depends on the Mach number
is not clear.
If high density regions are formed mainly through shock compression 
in a system with the rms Mach number ${\cal M}_{\rm rms}$, 
the average density contrast
$\langle \delta \rho/\rho \rangle \sim \langle \rho \rangle_V/\rho_0 -1$ 
could be described by ${\cal M}_{\rm rms}$, and it is expected that
$\langle \rho \rangle_V/\rho_0 = e^{\sigma^2/2} = 1 + {\cal M}_{\rm rms}^2$ 
using eq. (\ref{eq: msigma}). 
However, the ISM is not isothermal on a global scale.
An inhomogeneous galactic disk is characterized by a fully developed 
turbulence, and its velocity dispersion is a function of scales
as shown by power-law energy spectra \citep{wad02b}. 
This means that the galactic disk cannot be modeled with a single
`sound velocity' or velocity dispersion of the gas.
Therefore we cannot use the empirical relation on $\sigma$ for
galactic disks.

One should note that there is another problem on 
the argument based on
the rms-Mach number, which is in terms of origin and 
mechanism of maintaining the turbulence in molecular clouds.
Numerical experiments suggested 
that the turbulence in molecular clouds is dissipated in a sound
crossing time \citep{maclow99,ostr01}, and
there is no confirmed prescriptions on energy sources 
to compensate the dissipation.
Therefore, it is more natural to assume that the velocity dispersion 
in a molecular cloud is not constant in a galactic 
rotational period, and as a result the structure of
the density field is no longer static. If this is the case, 
taking the rms-Mach number as a major parameter to describe 
the global star formation rate would not be adequate.
The decaying turbulence could be supported by 
energy input by supernovae or out-flows by proto-stars, but 
even in that case there is no clear reason to assume
a uniform and time-independent Mach number.


\subsection{Observational evidence for LN-PDF}

It is practically difficult to know the PDF of the ISM in galactic
disks directly from observations.
We have to map the ISM 
in external galaxies with fine enough spatial resolution 
by observational probes that cover a wide density range.
The Atacama Large Millimeter/Submillimeter Array will be
an ideal tool for such observations.
Nevertheless, there is indirect evidence to support the LN-PDF in
the Large Magellanic Cloud (LMC), which is 
mapped by HI with a 15 pc resolution \citep{kim03}
 and by CO ($J=1-0$) with 8 pc resolution \citep{fukui01,yama01}. 
We found that the distribution function of
HI intensity distribution and a mass spectrum of CO clouds are consistent with
a numerical model of the LMC, in which
the density PDF in the simulation is
nicely fitted by a single log-normal function \citep{wad00}.
  Although we need more information about density field
by other probes, this suggests that the entire density filed of
the ISM in LMC could be modeled by a LN-PDF.

Recently, \citet{tozzi06} claimed that distribution of absorption column density
$N_{\rm H}$ in 82 X-ray bright sources in the Chandra Deep Field South
is well fitted by a log-normal function. This suggests that
the obscuring material around the AGNs is inhomogeneous as
suggested by previous numerical simulations \citep{wad02, wad05}, and their
density field is log-normal, 
if orientation of the obscuring ``torus'' is randomly 
distributed for the line of sight in the samples.

\subsection{The $Q$ criterion for star formation}
The origin of the Schmidt-Kennicutt relation 
on the star formation rate in galaxies
\citep{ken98} has been often discussed in terms of gravitational instability in 
galactic disks. More specifically
it is claimed that 
the threshold density for star formation can be represented by
the density for which the Toomre Q parameter is unity, 
i.e. $Q\equiv \kappa c_s/(\pi G \Sigma_g) = 1$,
 where $\kappa, c_s,$ and $\Sigma_g$ are epicyclic frequency, sound velocity, and surface density of gas \citep{ken98, mart01}.
However, this is not supported by recent observations 
in some gas-rich spiral galaxies \citep{wong02,koda05}.
Based on our picture described in this paper, it is not surprising that 
observationally determined $Q$  or the critical density 
do not correlate with the observed star formation rate.
Stars are formed in dense molecular clouds, which is 
gravitationally unstable on a {\it local} scale (e.g. $\lesssim$ 1 pc),
 but this is basically 
independent of the global stability of the galactic disk.
Even if a galactic gas disk is globally stable (e.g. effective $Q > 1$), cold, dense
molecular clouds should exist \cite[see also][]{wad99}.
Numerical results show that once the galactic disk is gravitationally and thermally 
unstable, inhomogeneous structures are developed, and in a non-linear phase,
it becomes ``globally stable'', in which the ISM is
turbulent and multi-phase, and its density field is characterized by LN-PDF.
 In that regime, SFR is determined by a fraction of dense clouds and
by a critical density for ``local star formation'', which should be
related to physical/chemical conditions in molecular clouds, not in
the galactic disk. In this picture, SFR naturally drops for 
less massive system without introducing critical density (see Figs. \ref{fig: sfr-vol} and \ref{fig: sfr}).

Another important point, but it has been often ignored, is
that the $Q$ criterion is derived from a dispersion relation 
for a {\it tightly wound} spiral perturbation in a 
thin disk with a uniform density \cite[see e.g.][]{binney87}.
The equation of state is simply assumed as $p \propto \Sigma^\gamma$, where
$\Sigma$ is surface density and $\gamma$ is a constant.
The criterion for instability, i.e. $Q=1$ means that the disk is 
{\it linearly} unstable for an {\it axisymmetric} perturbation.
These assumptions are far from describing star formation criterion 
in an inhomogeneous, multi-phase galactic disks.
One should also note that
gas disks could be unstable for non-axisymmetric perturbations, even 
if $Q > 1$ \citep{gold65}.

Finally, we should emphasize that 
it is  observationally difficult to determine $Q$ precisely.
All the three variables in the definition of $Q$, 
i.e. surface density of the gas, epicyclic frequency, and sound velocity,
are not intrinsically free from large observational errors.
Especially, it is not straightforward to define the `sound velocity', $c_s$ in
the multi-phase, inhomogeneous medium, and
there is no reliable way to determine $c_s$ by observations.
Therefore one should be careful for discussion
based on the {\it absolute} value of $Q$ in terms of star formation criterion.


\subsection{LN-PDF and star formation rate in simulations}

\citet{krav05} run a cosmological N-body/AMR simulation to
study formation of globular clusters in a Milky Way-size galaxy. 
They found that the density PDF in the galaxy
can be fitted by a log-normal function, and it evolves with
the redshift. The dispersion for the exponent of a natural log is
$\sigma \simeq 1.3, 2.0,$ and 2.8 at $z=7, 3.3$ and 0, respectively.
As shown in the previous section, the dispersion of the LN-PDF
is related with the average gas density in the system.
Therefore, the increase of the dispersion in the galaxy formation 
simulation should be because of increase of the gas density and/or
decrease of the characteristic density $\arho$. 
The former can be caused by accretion of the gas, in fact,
\citet{krav05} shows that the gas density increases until $z \simeq 5$.
They also shows that the characteristic density $\arho$ decreases from $z \simeq 7$ to $z \simeq 3$. 
Based on the argument on $p_{\rm eff}$ in \S 3.1, this is reasonable, 
if 1) a clumpy, turbulent structure is developed due to gravitational instability
and 2) temperature of the lower density gas decreases due to
radiative cooling with time.
Qualitatively this is expected because at lower redshift, heating due to
star formation and shocks caused by mergers are less effective,
but radiative cooling becomes more efficient due to increase of metallicity
and gas density.
\citet{krav05} mentioned that widening the LN-PDF with
decreasing redshift is due to increase of the rms Mach number of 
gas clouds. However, we suspect that this is not the case in formation of galactic
disks (see discussion in \S 4.1).

\citet{li05} investigated star formation in an isolated galaxy, using
three-dimensional SPH code (GADGET). They use sink particles to directly measure the mass of gravitationally collapsing gas, a part of which is considered as newly formed 
stars.
They claimed that the Schmidt law observed in disk galaxies is quantitatively reproduced.
They suggest that the non-linear development of gravitational instability determines the
local and global Schmidt laws and the star formation rate.
Their model SFR shows a rapid decline for decreasing the surface density, which is
consistent with our prediction (Fig. 11).

\subsection{Effects of energy feedback}

\begin{figure}[t]
\centering
\includegraphics[width = 8cm]{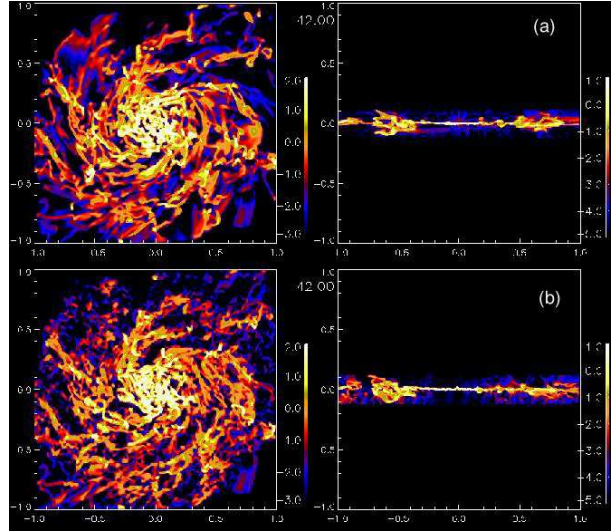}
\caption{Same as Fig. 1, but comparison between models with and without energy feed back from supernovae. (a) model D, (b) model D, but with supernova explosions at a rate of 1.5$\times 10^{-3}$ yr$^{-1}$ kpc$^{-2}$.
}
\label{fig: snmodel}
\end{figure}

\begin{figure}[t]
\centering
\includegraphics[width = 8 cm]{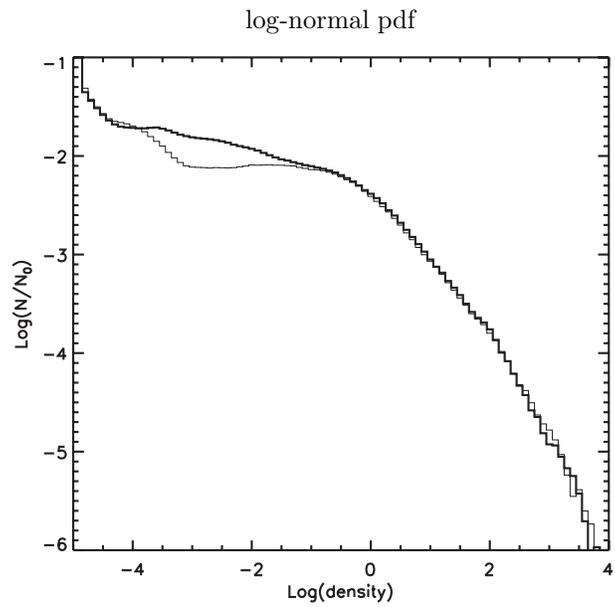}
\caption{PDF in models with and without energy feedback from supernovae. Thin line is  model D, and thick line is model D, but with supernova explosions at a rate of 1.5$\times 10^{-3}$ yr$^{-1}$ kpc$^{-2}$.
}
\label{fig: pdf_snmodel}
\end{figure}

In \S 2, we focus on models without energy feedback from supernovae
in order to know an intrinsic structures of the ISM, which is dominated by
gravitational and thermal instabilities.
However, one should note 
that the LN-PDF is robust for including energy feedback from supernovae 
as suggested by previous 2-D models for galactic disks and 3-D models for 
galactic central regions (WN01 and W01).
This is reasonable, because stochastic explosions in
the inhomogeneous medium is a preferable situation for the LN-PDF.
In order to confirm this in 3-D on a galactic scale, we run a model with
energy feedback from supernovae, in which a relatively large supernova rate 
($1.5\times 10^{-3}$ yr$^{-1}$ kpc$^{-2}$) is assumed.
Energy from a supernova ($10^{51}$ ergs) is injected into one grid cell
which is randomly selected in the disk.
In Figs. \ref{fig: snmodel} and \ref{fig: pdf_snmodel},  we show density distributions
and PDFs in models with and without energy feedback. 
It is clear that even for the large supernova rate the density distribution and
PDF are not significantly different from the model without energy feedback, especially
for the regime above the characteristic density.

\subsection{Origin of a bias in galaxy formation}

It is observationally suggested that massive galaxies terminate their 
star formation at higher redshift than less massive galaxies, namely "down-sizing" in
galaxy formation \citep[e.g.][]{kauf03, kodama04}. This seems to be 
inconsistent with a standard hierarchical clustering scenario. It is a puzzle why 
the star formation time scale is shorter in more massive
galaxies, in other words, why the star formation 
rate and/or star formation efficiency are extremely biased in an 
environment to form massive galaxies. This might be understood by our result
 of the global star formation,
that is  SFR increases extensively as a function of average gas density 
(Fig. 11),
e.g. SFR $\propto \rho^2$ ($\rho \sim 1 M_\odot$ pc$^{-3}$, for $\delta_c 
= 10^4$).
Since the average baryon density is higher in an environment where 
massive 
galaxies could be formed, SFR could be 100 times larger, if the 
average gas density
is 10 times higher. It would be interesting to note 
that this tendency is more extreme for higher 
critical density ($\rho_c$). Critical density for star formation 
can be affected by UV radiation through
 photoevapolation of dense clouds 
in proto-galactic halos \citep[e.g.][]{susa04}. Qualitatively this means that 
for a stronger UV field, 
the critical density is larger, therefore SFR can strongly depend
on the average gas density.






%
\section{CONCLUSION}
%

Three-dimensional, high resolution hydrodynamic simulations of 
a galactic disk shows that the density probability distribution function (PDF)
is well fitted by a single log-normal function over 6 decades.
The dispersion of the log-normal PDF (LN-PDF), $\sigma$ can be described by
$
\sigma = {1}/{2} \ln \left({\langle\rho\rangle_M}/{\rho_0}\right),
$
where $\arho$ is a characteristic density,
and $\langle\rho\rangle_M$ is mass-weighted average density of the ISM (eq. (\ref{eqn: eq17b})).
If all the ISM is in the regime of LN-PDF, the star formation rate
can be represented by eq. (\ref{eq: modelsfr}).
We find that star formation rate (SFR) is sensitive for increasing
average gas density, especially for smaller $\langle\rho\rangle_V$
and larger $\rho_c$. It is however that SFR does not significantly depend
on the characteristic density $\rho_0$. 
We compare the observed star formation rate 
in normal and spiral galaxies, and find that 
a model  with $\rho_c \simeq 10^3$ cm$^{-3}$,
a volume fraction of the LN-part $\simeq 0.1$, and $\rho_0 \simeq 1$ cm$^{-3}$,
well explains the observed trend of SFR as a function of
average surface density.
If the scale height of the ISM in star forming regions is
$\simeq 100$ pc, the star formation efficiency in starburst galaxies 
is 0.1-0.01, and it is one-order smaller in normal galaxies.

The log-normal nature of the density field should be intrinsic
in an inhomogeneous, multi-phase ISM, 
if 1) the whole system is globally quasi-stable in a long enough period
(for a galactic disk, it is at least a few rotational periods), 
2) the system is consisted of many hierarchical
sub-structures, and 3) density in such sub-structure is determined by 
random, non-linear, and independent processes. 
If these conditions are satisfied, 
any random and nonlinear processes that affect a density field
should cause the log-normal probability distribution function.
In this sense, most physical processes expected in real galactic disks,
such as nonlinear development of 
magneto-rotational instability, interactions between the ISM and
stellar wind, and heating due to non-uniform radiation fields 
originated in OB associations should also generate the log-normal PDF.
These effects on the PDF could be verified in more realistic
numerical simulations with a wide dynamic range in the near future.
Observational verification on the density PDF of the ISM in 
various phases is also desirable.

%
\acknowledgments 
%
The authors are grateful to the anonymous referee for his/her careful reviewing and 
constructive comments.
Numerical computations were carried out on Fujitsu VPP5000 at NAOJ.  
KW is supported by Grant-in-Aids for Scientific Research 
[no. 15684003 and 16204012 (KW)] of JSPS.
\newpage

\clearpage

\clearpage

\clearpage
\end{document}